\documentclass[reprint,aps,pra,superscriptaddress]{revtex4-1}
\usepackage{braket,mathtools,amssymb,bm,amsmath}
\usepackage[utf8]{inputenc}
\pdfoutput=1

\begin{document}

\title{Nonclassical Kernels in Continuous Variable Systems}
\author{Roohollah Ghobadi}
\thanks{farid.ghobadi80@gmail.com}
\affiliation{Institute for Quantum Science and Technology and
Department of Physics and Astronomy, University of Calgary, Calgary, Alberta,
Canada T2N 1N4}
\date{\today}

\begin{abstract}
Kernel methods are ubiquitous in classical machine learning, and recently their formal similarity with quantum mechanics has been established. To grasp the potential advantage of quantum machine learning, it is necessary to understand the distinction between non-classical kernel functions and classical kernels. This paper builds on a recently proposed phase space nonclassicality witness [Bohmann, Agudelo, Phys. Rev. Lett. 124, 133601 (2020)] to derive a witness for the kernel's quantumness in continuous-variable systems. We discuss the role of kernel's nonclassicality in data distribution in the feature space and the effect of imperfect state preparation. Furthermore, we show that the non-classical kernels lead to the quantum advantage in parameter estimation. Our work highlights the role of the phase space correlation functions in understanding the distinction between classical machine learning from quantum machine learning. 

\end{abstract}
\maketitle
\date{\today}

Small scale quantum systems are a new testbed for understanding the potential of quantum enabled technologies~\cite{Google,Zhong,Xanadu,RL}. A promising application of the small-scale quantum systems is in machine learning, where the goal is to reveal patterns in data~\cite{biamonte2017quantum, RMP19} or designing intelligent agents for reinforced learning~\cite{RL}. One scenario for taking advantage of these intermediate devices in machine learning is using them to estimate a classically expensive kernel function, a similarity measure between data points. A classical computer can then find the best performing hypothesis~\cite{schuld2019quantum,NatureIBM}. 

The quantum kernel methods provide a framework for understanding the quantum advantage in machine learning. For example, the generalization performance in supervised machine learning, which quantifies how accurate a learning model can classify unseen data, has recently been studied in this framework ~\cite{Huang2}; and a necessary condition for achieving a quantum advantage in generalization has been derived~\cite{Huang2}. These works highlight the importance of understanding the difference between quantum and classical kernel functions to assess the quantum advantage in machine learning.  
 
Photonic hardware is a promising path for implementing and scaling up the quantum algorithms~\cite{Bourassa}. Some groundbreaking demonstrations such as Gaussian boson sampling~\cite{RL,Xanadu}, photonic neural network~\cite{Steinbrecher}, quantum-enhanced data classification~\cite{Xia} and quantum advantage in reinforcement learning~\cite{RL} have recently been reported based on photonic hardware. From a theoretical standpoint, an appealing property of quantum optical platforms compare to qubit systems is their well-understood phase space structures. At the conceptual level, this makes it possible to distinguish the genuine quantum effects from classical ones using quasiprobability distributions \cite{Cahill,Spekkens,Sperling} such as the Wigner function and to identify the enabling resource for quantum computing \cite{Howard}. 

A crucial step in any quantum machine learning algorithm involves encoding the data in quantum states \cite{Schuld21,Schuld212}. The data could be either data obtained from sampling a classical system or data acquired from quantum measurements. Recently Xia \textit{et al.} demonstrated the quantum advantage in supervised machine learning in a setting where an entangled sensor network provides the input data \cite{Xia}. This approach relies on the well-understood quantum advantage in metrology and allows for applying the tools from quantum sensing for machine-learning tasks.    

In this paper, building on a recently proposed phase-space inequality \cite{BohmanPRL, BohmanQuantum, Biagi}, we drive a nonclassicality witness for kernel functions. We discuss the role of kernel's nonclassicality in data distribution in feature space. The consequence of the imperfect state preparation is considered. We show the quantum advantage of nonclassical kernels for sensing and parameter estimation. The edge of nonclassical kernels in parameter estimation suggests the benefit of employing data provided from the quantum sensor for supervised learning assisted \cite{Xia,Zhuang}. Since the nonclassicality of a quantum state can be converted to quantum entanglement \cite{Kim,Killoran}, our results can be seen as an indirect study of the role of quantum correlation in machine learning \cite{Xia,Zhuang}.

{\it Supervised machine learning.} Given a labeled dataset $\{(x_{k},f(x_{k})), x_{k}\in\mathbb{R}^{N},k=1,...M\}$, the task is to find a mapping $f(x)$ for new (unlabeled) data point. In the case of classification $f(x_{k})=\pm1$ and for regression $f$ is a continuous function of $x$. The data set can be either classical data available a priori or data collected from some quantum measurements. Formally the learned function $h^{*}(x)$ is the solution of the following equation \cite{scholkopf2001generalized,hofmann2008kernel}

\begin{equation}\label{model}
h^{*}(x)=\text{arg min}_{h\in\mathcal{H}}\frac{1}{M}\sum_{i=1}^{M}\mathcal{L}(h(x_{i}),y_{i}) +\lambda||h||^{2},
\end{equation}

where $\mathcal{L}$ is a loss function that characterizes the performance of the learned function, and the $\mathcal{H}$ is the space of hypothesis. The second term is known as the regularization term with $\lambda$ as the regularization coefficient. The regularization term in Eq.(\ref{model}) penalizes complex hypotheses that best match the training data and favors a smooth one with better prediction performance. 

The kernel method in machine learning is based on mapping(encoding) data into a Hilbert space by the feature map $\Phi:\mathcal{X}\rightarrow \mathcal{H}$ where $\mathcal{H}$ is a Hilbert space. One can then use the inner product $\langle.,.\rangle_{\mathcal{H}}$ over $\mathcal{H}$ to define the kernel function  \cite{scholkopf2001generalized,hofmann2008kernel}

\begin{equation}\label{kerneldef0}
K(x,x')=\langle\Phi(x),\Phi(x')\rangle_{\mathcal{H}},
\end{equation}

a measure of similarity between for $x,x'\in\mathcal{X}$. The kernel is symmetric $K(x,x')=K(x',x)$ and positive semidefinite $\forall x_{i},x_{j}\in\mathcal{X}$, {\it i.e.}  $\sum_{i,j}c_{i}c_{j}K(x_{i},x_{j})\geq0$. A useful family of hypothesis, known as Reproducing kernel Hilbert space (RKHS), can be constructed from a positive kernel function as \cite{scholkopf2001generalized,hofmann2008kernel}

\begin{equation}
\mathcal{R}_{K}=\{h:
\mathcal{X}\rightarrow\mathbb{C}|h(x)=\langle w,\Phi(x)\rangle_{\mathcal{H}},w,\Phi\in\mathcal{H}\},
\end{equation}

with the reproducing property $h(x)=\langle h,K_{x}\rangle_{\mathcal{H}}$ for $h\in\mathcal{H}$. Therefore $K_{x}\in\mathcal{H}$ is the evaluation functional which maps $h$ to $h(x)$. The advantage of RKHS is best shown using the representer theorem that grantees that the solution of Eq.(\ref{model}) in the RKHS is given by \cite{scholkopf2001generalized,hofmann2008kernel}

\begin{equation}\label{Repsenter}
h^{*}(x)=\sum_{x_{k}\in\mathcal{X}}c_{k}K(x,x_{k}),
\end{equation}

with $c_{k}\in\mathbb{R}$ to be determined. For the square loss function $\mathcal{L}(h(x_{i}),y_{i})=(h(x_{i})-y_{i})^{2}$ from Eqs.(\ref{model},\ref{Repsenter}) one obtains $C=(K+\lambda M\mathbb{I})^{-1}Y$.

{\it Generalization error bound.} For a class of bounded functions  $\{ f \in\mathcal{R}_{K}:||f||\leq B\}$ the generalization error bound is given by \cite{kernelbook}

\begin{equation}\label{error}
\mathbb{E}_{x\in\mathcal{D}}|h^{*}(x)-f(x)|\leq \frac{2B}{\sqrt{M}},
\end{equation}

where the average is with respect to a fixed distribution $\mathcal{D}$ over input data 
and $B=||f||^{2}=C^{T}KC=Y^{T}(K+M\lambda\mathbb{I})^{-1}K(K+M\lambda\mathbb{I})^{-1}Y$. From Eq.(\ref{error}), one can see that the kernel function controls the generalization error bound thorough $B$. 

{\it Quantum machine learning.} A remarkable similarity between kernel methods and quantum mechanics is revealed by defining the feature map as $x\rightarrow\rho(x)$ with $\rho$ as a quantum state in the Hilbert space of the quantum system \cite{schuld2019quantum, NatureIBM}. The mapping from data to quantum states is understood as the action of unitary
operator $U$ as

\begin{equation}\label{encoding}
\rho(x)=U^{\dagger}(x)\rho U(x). 
\end{equation}

The kernel function is then defined as 

\begin{equation}\label{kernelQ}
K_{\rho}(x,x')=\text{tr}(\rho(x)\rho(x')).
\end{equation}


{\it Nonclassicality in phase space.} The phase-space approach to quantum mechanics would allow distinguishing the quantum mechanics from the classical statistical mechanics. To see this consider the coherent state which is defined as $|\alpha\rangle=D(\alpha)|0\rangle$ where $D(\alpha)=e^{\alpha a^{\dagger}-\alpha^{*}a}$ is the displacement operator and $|0\rangle$ is the vacuum state and $a$ and $a^{\dagger}$ as annihilation and creation operator, respectively. We can expand arbitrary quantum state in the basis of the coherent state as \cite{Glauber,Sudarshan}

\begin{equation}\label{Glauber}
\rho=\int d^{2}\alpha P(\alpha)|\alpha\rangle\langle\alpha|,
\end{equation}

where $P(\alpha)$ is known as the Glauber-Sudarshan $P$ distribution. In the case of classical state $ P(\alpha)$ is a valid (density) probability distribution {\it i.e.} $P(\alpha)\geq0$. Accordingly, violation of $P(\alpha)\geq0$ signals the nonclassical nature of state states. Instead of working with $P$ distribution, it is more convenient to work with quasiprobability distributions such as the Wigner function, which are non-singular and experimentally accessible \cite{Leonhardt}. In this paper, we are especially interested in the following inequality, which holds for classical states \cite{BohmanPRL, BohmanQuantum}

\begin{equation}\label{witnessW}
W_{\rho}(\bm{\alpha})W_{\rho}(\bm{\alpha}')-e^{-|\bm{\alpha}-\bm{\alpha'}|^{2}}\\ W_{\rho}(\frac{\bm{\alpha}+\bm{\alpha'}}{2})^{2}\geq0,
\end{equation}

where $W_{\rho}$ is the Wigner function of density matrix $\rho$, $\bm{\alpha}=(\alpha_{1},...,\alpha_{N})$ are arbitrary points in the phase space with $\alpha_{i}=\frac{q_{i}+p_{i}}{\sqrt{2}}$. The violation of inequality Eq. (\ref{witnessW}) for $\bm{\alpha}\neq\bm{\alpha'}$ signals the nonclassicality of the state. The equality in Eq.(\ref{witnessW}) is attained by coherent states. 
Notably, (\ref{witnessW}) can capture the quantum nature of states with non-negative Wigner functions \cite{BohmanPRL, BohmanQuantum}. 

The effect of encoding map $U(x)$ in Eq.(\ref{encoding}) can be best understood by its action on the annihilating operator $a\rightarrow U(x)a U^{\dagger}(x)$. Equivalently, in the phase space picture this mapping translated to $\bm{\alpha}\rightarrow \langle U(x)a U^{\dagger}(x)\rangle:=\bm{\alpha}_{x}$. Next we use the covariance property of the Wigner function \cite{Gibbons} to write $W_{\rho}(\bm{\alpha}_{x})=W_{\rho(x)}(\bm{\alpha})$. Therefore applying the encoding  Eq.(\ref{encoding}) to the non-classicality condition (\ref{witnessW}), one obtains  

\begin{equation}\label{step}
W_{\rho(x)}(\bm{\alpha})W_{\rho(x')}(\bm{\alpha})-e^{-|\bm{\alpha}_{x}-\bm{\alpha'}_{x}|^{2}}W_{\rho}^{2}(\frac{\bm{\alpha}_{x}+\bm{\alpha}_{x'}}{2})\geq0.
\end{equation}

Next we integrate over entire phase space and use the overlap formula \cite{Leonhardt} $\text{tr}(AB)=2\pi\int d^{2}\bm{\alpha} W_{A}(\bm{\alpha})W_{B}(\bm{\alpha})$ 
where $W_{A}$ and $W_{B}$ are the Wigner functions of operators $A$ and $B$, respectively, and use Eq.(\ref{kernelQ}) to obtain

\begin{equation}\label{gencon}
K_{\rho}(x,x')\geq \int d^{2N}\bm{\alpha}e^{-|\bm{\alpha}_{x}-\bm{\alpha'}_{x}|^{2}}W_{\rho}^{2}(\frac{\bm{\alpha}_{x}+\bm{\alpha}_{x'}}{2}).
\end{equation}

On the virtue of the Eq.(\ref{witnessW}), we call a kernel to be a nonclassical kernel only if (\ref{gencon}) is violated. One can define
the kernel’s nonclassicality witness 

\begin{equation}\label{gencon1}
\mathcal{W}_{\rho}(x,x'):=K_{\rho}(x,x')-\int d^{2N}\bm{\alpha}e^{|\bm{\alpha}_{x}-\bm{\alpha'}_{x}|^{2}}W_{\rho}^{2}(\frac{\bm{\alpha}_x+\bm{\alpha}_x'}{2}),
\end{equation}

which is always positive for classical kernels, while it can become negative for nonclassical kernels.

A simple version of (\ref{gencon}) can be obtain if we choose to encode data points using the  displacement operator $U(x)=D(x)$ for $x\in\mathbb{C}$ (see also \cite{Chatterjee}). In this case, upon using $\bm{\alpha}_{x}=\bm{\alpha}-x$ and $\bm{\alpha}_{x'}=\bm{\alpha}-x'$ in (\ref{gencon}) one gets  

\begin{equation}\label{Kernelphase2}
K_{\rho}(x,x')\geq e^{-|x-x'|^{2}}\text{tr}\rho^{2}.
\end{equation}

{\it Parameter estimation.} It is interesting to note that the conditions (\ref{gencon},\ref{gencon1}) can be applied to the parameter estimation. In a typical quantum metrology experiment, a quantum state $\rho$ known as probe interacts with a specific degree of freedom of a physical system that we are interested in measuring. By reading out the change in the probe state, one can estimate the system's parameter of interest. Formally, the interaction of the probe-system is described by a unitary operator $U(x)$ with $x$ as an unknown parameter to be estimated. As a result of this interaction the state of probe evolved to $\rho(x)=U(x)\rho U^{\dagger}(x)$. A figure of merit that captures the precision of the estimation is the relative purity \cite{delCampo} 

\begin{equation}\label{metric}
f_{\rho}(x,x'):=\frac{\text{tr}(\rho(x)\rho(x'))}{\text{tr}\rho^{2}},
\end{equation}

which is a measure of sensitivity to distinguish quantum states $\rho(x),\rho(x')$: the smaller $f_{\rho}(x,x')$ is, the easier it would be to discriminate between quantum states $\rho(x),\rho(x')$.  Interpreting the data encoding as given by Eq.(\ref{encoding}) as data imprinting into the probe state, and comparing Eq.(\ref{metric}) with Eq.(\ref{kernelQ}), it is clear that up to a normalization factor, the relative purity is identical to the kernel function. Therefore one can use (\ref{gencon1}) to define the nonclassicality witness for parameter estimation as 

\begin{equation}\label{pest}
f_{\rho}(x,x')\geq\frac{1}{\text{tr}\rho^{2}}\int d^{2N}\bm{\alpha}e^{-|\bm{\alpha}_{x}-\bm{\alpha'}_{x}|^{2}}W_{\rho}^{2}(\frac{\bm{\alpha}_{x}+\bm{\alpha}_{x'}}{2}),
\end{equation}

with the lower bound achieved for coherent states. A violation of condition (\ref{pest}) is a signature of quantum advantage in parameter estimation. 

If the probe state undergoes a displacement proportional to the signal’s strength, from Eqs.(\ref{Kernelphase2},\ref{pest}) one gets $f_{\rho}(x,x')\geq e^{-|x-x'|^{2}}$. Reminding that $e^{-|x-x'|^{2}}$ is the kernel of coherent state written in the shot noise unit, the nonclassicality in the parameter estimation reduces to beating the shot noise limit {\it i.e.} $\Delta x_{\rho}=\frac{1}{\sqrt{\partial_{x}^{2}f_{\rho}}}<\Delta x_{c}=1$ with $\Delta x_{\rho}$ and $\Delta x_{c}$ as variance of quantum state $\rho$ and coherent state respectively. 

The nonclassicality of quantum state can be converted to the quantum entanglement using a simple linear optical circuits with beam splitter and phase shifters \cite{Kim,Killoran}. Denoting the effect of linear circuit by $U_{L}$, including the linear circuit amounts to replacing $U(x)$ with $U_{L}U(x)$ in Eq.(\ref{encoding}) and $\bm{\alpha}\rightarrow\langle U_{L}U(x)aU^{\dagger}(x)U^{\dagger}_{L}\rangle:=\bm{\alpha}_{x}$ in Eq.(\ref{step}).

{\it Data distribution in feature space.} The kernel function in Eq.(\ref{kernelQ}) determines the induced distance between data points in the Hilbert space. For two arbitrary data points  $x,x'\in\mathcal{X}$ (or signals measured by some quantum measurements) the distance between their images $\rho(x)$ and $\rho(x')$ reads $d_{\rho}^{2}(x,x')=||\rho(x)-\rho(x')||^{2}=\langle \rho(x)-\rho(x'),\rho(x)-\rho(x')\rangle$ or equivalently

\begin{equation}\label{distance}
d_{\rho}^{2}(x,x')=2(\text{tr}\rho^{2}-K_{\rho}(x,x')),
\end{equation}

where we used $K(x,x)=K(x',x')=\text{tr}\rho^{2}$ for the unitary encoding. Since $d_{\rho}^{2}(x,x')\geq0$ it follows that $K_{\rho}(x,x')\leq\text{tr}\rho^{2}$. The $\text{tr}\rho^{2}$ is known as the purity of the state which takes values between zero and one, corresponding to completely mixed and pure states, respectively. From Eq.(\ref{distance}), we conclude that the effect of imperfect state preparation, as captured by decreasing the purity of the state, is to reduce the distance between the data points in the feature space.

To benchmark the performance of nonclassical state, we write Eq.(\ref{distance}) for coherent states

\begin{equation}\label{discal}
d_{\text{c}}^{2}(x,x')=2(1-K_{c}(x,x')), 
\end{equation}

where $K_{c}(x,x')$ denotes the kernel of the coherent state $\rho_{c}$. Applying Eq.(\ref{gencon}) in Eq.(\ref{discal}) one obtains

\begin{equation}\label{distanceQ0}
d_{\rho}(x,x')>d_{c}(x,x'). 
\end{equation} 

Therefore a nonclassical kernel could lead to a better separation between data points. The quantum advantage in separating data points survives for as long as $K_{c}(x,x')-K_{\rho}(x,x')\geq1-\text{tr}\rho^{2}$.

Next, defining the center of mass of a data set by $\bar{\rho}=\frac{1}{M}\sum_{i=1}^{M}\rho(x_{i})$, from (\ref{Kernelphase2}) it follows that $||\bar{\rho}_{c}||^{2}>||\bar{\rho}||^{2}$ with $||\rho(x)||^{2}=\langle\rho(x),\rho(x)\rangle$. As a consequence nonclassical kernels results in a greater average distance from the center of mass of training data. Another observation is that 

\begin{equation}\label{avedis}
\frac{1}{M}\sum_{i=1}^{M}||\rho(x_{i})-\bar{\rho}||^{2}>\frac{1}{M}\sum_{i=1}^{M}||\rho_{c}(x_{i})-\bar{\rho}_{c}||^{2},
\end{equation}

therefore nonclassical kernel results in a greater distance from the center of mass. The improvement in clustering and classification based on data gathered from quantum sensors has been observed experimentally in \cite{Xia}.

{\it Examples.} In the following, we give a few examples for quantum kernel kernels based on the displacement encoding and their corresponding witness based on condition (\ref{gencon1}). In all of these examples, we use a single-mode quantum state for simplicity.

{\it Squeezed state.} The squeezed state is defined as $|S\rangle=S(r)|0\rangle$ where
$S(r)=e^{r(a^{\dagger 2}-a^{2})}$ is the squeezing operator and $r$ is squeezing parameter. The action of squeezing operator in phase space can be understood by the transformations $q\rightarrow e^{r}q$ and $p\rightarrow e^{-r}p$. The Wigner function for squeezed state reads $W_{\text{sq}}(q,p)=\frac{1}{\pi}e^{-(e^{2r}q^{2}+e^{-2r}p^{2})}$. After some straightforward calculation, the kernel function of the damped squeezed state  is  

\begin{multline}\label{witnesssq}
K_{|S\rangle\langle S|}(x,x')=\text{tr}\rho^{2}(e^{-\frac{(x_{1}-x'_{1})^{2}}{1-\eta(1-e^{-2r})}}e^{-\frac{(x_{2}-x'_{2})^{2}}{1-\eta(1-e^{2r})}}),
\end{multline}

where $\text{tr}\rho^{2}=(1+4\eta(1-\eta)\sinh^{2} r)^{-1/2}$. 

{\it Single photon.} The single photon state is defined as $|1\rangle=a^{\dagger}|0\rangle$ with the Wigner function $W_{|1\rangle\langle1|}(q,p)=\frac{2(q^{2}+p^{2})-1}{\pi}e^{-\frac{q^{2}+p^{2}}{2}}$. The single photon state is a non Gaussian state which in contrast to squeezed state it can take negative values. The single photon kernel function then reads 

\begin{equation}\label{witnesss1}
K_{|1\rangle\langle1|}(x,x')=\eta|x-x'|^{2}e^{-|x-x'|^{2}}[2+\eta(|x-x'|^{2}-4)],
\end{equation}

and $\text{tr}\rho^{2}=(1-2\eta+2\eta^{2})$. 

\begin{figure}
\centering
\includegraphics[width=0.5\textwidth]{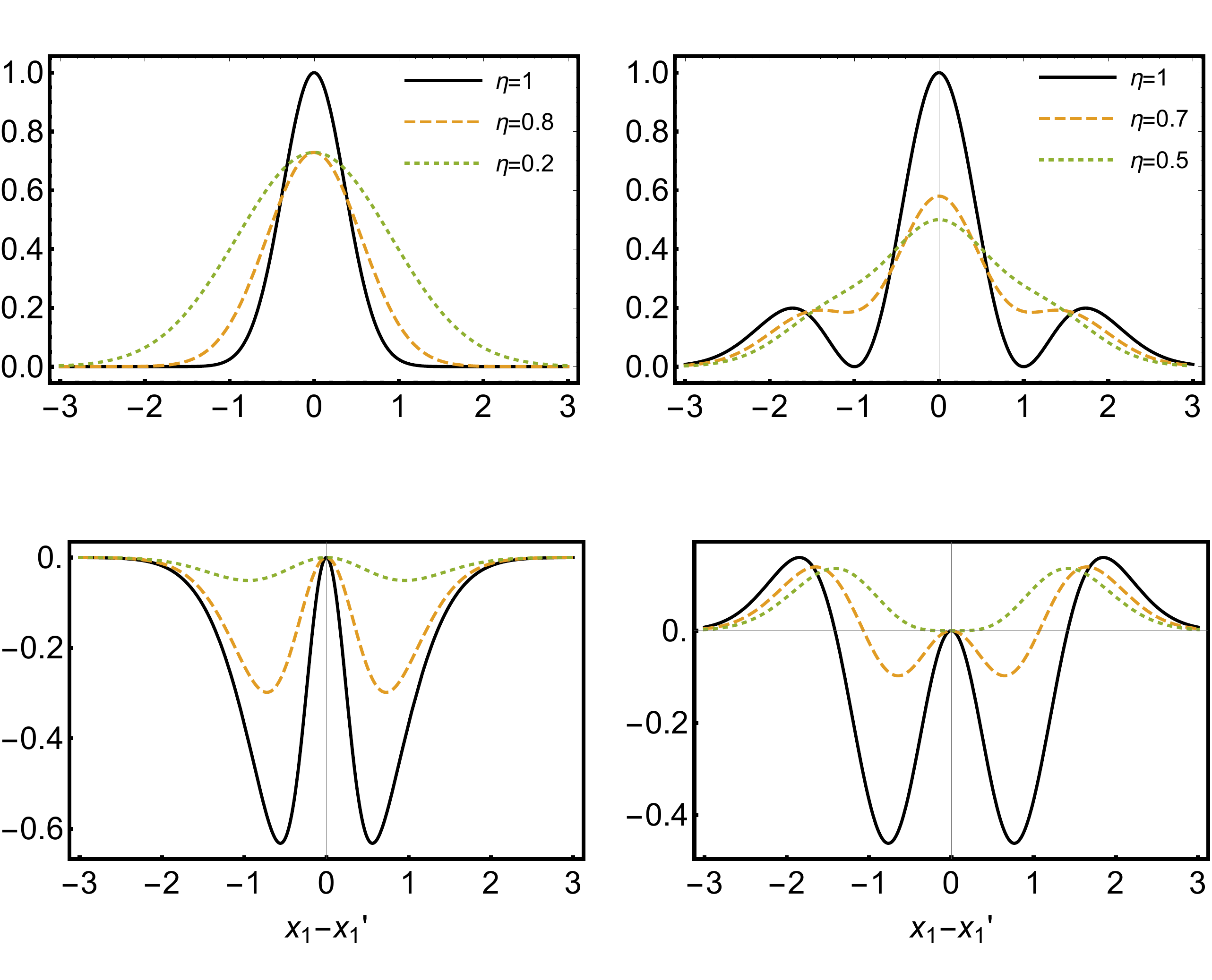}
\caption{The Kernel function for the squeezed state (upper left) and single photon state (upper right) for different loss as shown in the inset. In the lower row the non-classicality witness $\mathcal{W}$ for squeezed state (lower left) and single photon state (lower right) are shown. For the squeezing parameter we used $r=1$ and $x_{2}=x'_{2}=0$.}
\end{figure}

In Fig.(1) (upper row) we have plotted the kernel function for squeezed (we set $x_{2}=x'_{2}=0$ in Eq.(\ref{witnesssq})) and single-photon state, respectively, for different value of loss. As can be seen, the effect of loss for the squeezed state is to broaden the width of the kernel function and to shift the minimum of the $\mathcal{W}$ to the right; see also Eq.(\ref{witnesssq}). From Fig. (1), one can see the same trend for the single-photon state with the difference that the minimum of $\mathcal{W}$ shifts to the left. In Fig.(1) (lower row), the $\mathcal{W}(x,x')$ is shown for the squeezed state (lower left) and single-photon state (lower right). 

The squeezed state's kernel is quite robust against loss. To see this, we expand the Eq.(\ref{witnesssq}) around $\eta=0$ and $\mathcal{W}(x,x')=-\eta |x-x'|^{2}e^{-|x-x'|^{2}}(1-e^{-2r})$ which remains negative. On the other hand, doing the similar calculation for the single photon state's kernel shows that nonclassicality persists only for $\eta>50\%$. 

As a final example, we consider the parameter estimation based on Eq.(\ref{pest}). We consider case where the probe state undergoes displacement due to its interaction with the system. From Eq.(\ref{witnesssq}) and Eq.(\ref{witnesss1}) we obtain the variance of squeezed and single photon as $\Delta x_{|S\rangle\langle S|}=\sqrt{1-\eta(1-e^{-2r})}$ and $\Delta x_{|1\rangle\langle1|}=\sqrt{\frac{1+2\eta^{2}-2\eta}{1+6\eta^{2}-4\eta}}$. In Fig.(2) we show the variance of single photon, squeezed and coherent state as a function of transmissivity $\eta$. 

\vspace{0.4cm}

\begin{figure}[hbt!]\label{fig2}
\centering
\includegraphics[width=0.35\textwidth]{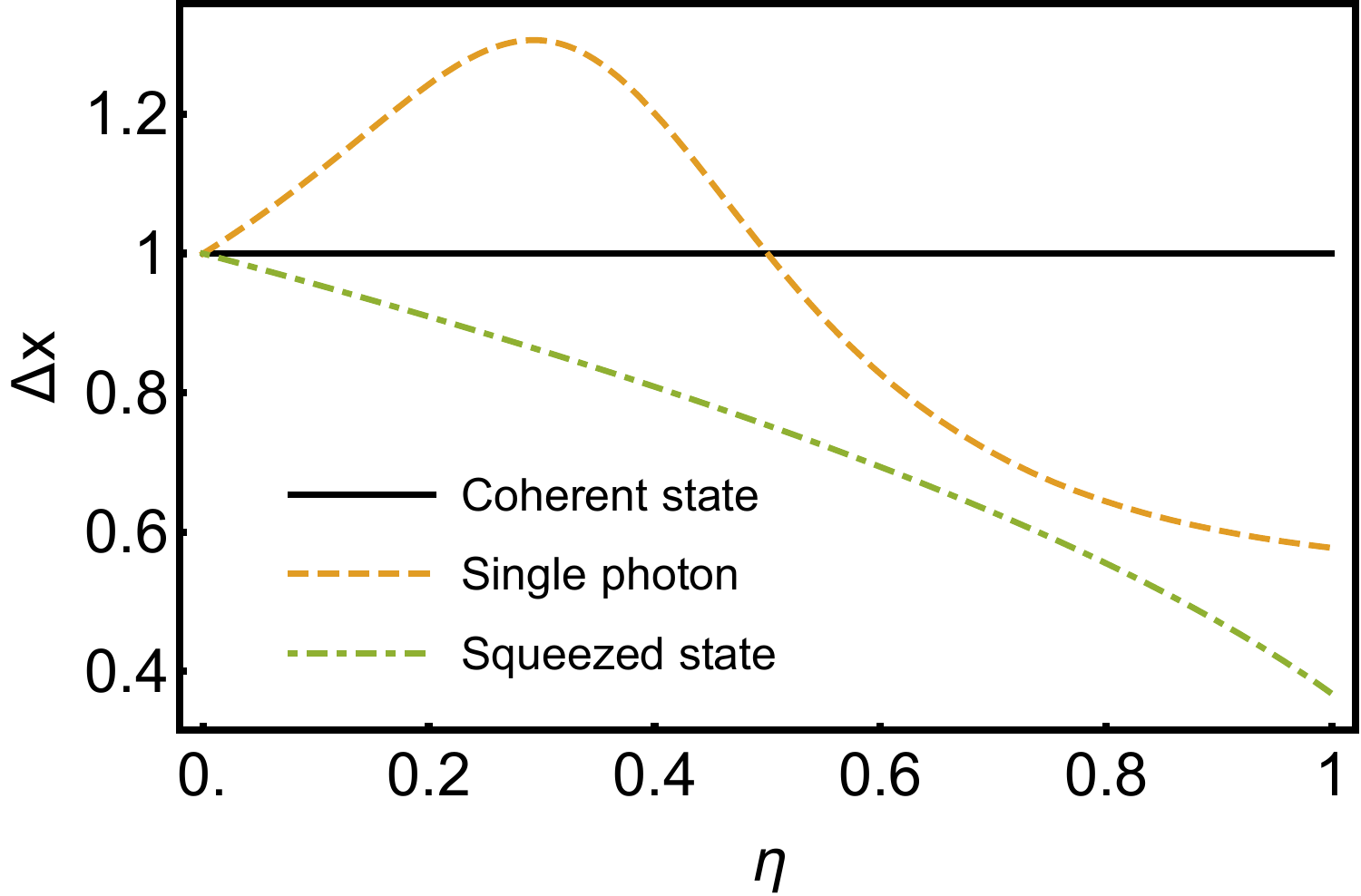}
\caption{The minimum detectable displacement using coherent state, squeezed state and single photon state. We used $r=1$ and $x_{2}=x'_{2}=0$.}
\end{figure}

We would like to emphasize that the kernel functions in Eqs. (\ref{witnesssq},\ref{witnesss1}) are easy to estimate classically and are shown as simple applications of (12-14). For classical data available a priori, it is necessary to consider the multimode states, which allow more complicated encoding, and the quantum entanglement can give rise to computationally expensive kernels \cite{Gu,schuld2019quantum}. Examples of circuits leading to computationally expensive kernels for qubit systems are given in \cite{NatureIBM, Liu}. For data acquired from quantum measurements, the quantum advantage in sensing can be enhanced by a factor of $\sqrt{M}$ where $M$ is the number of sensor nodes \cite{Xia}.


{\it Discussion and conclusion.} We identified the necessary condition for a kernel function to be nonclassical and its role in data distribution. The connection between kernel's nonclassicality and quantum enhanced sensing has been established. Our approach provides a simple picture of the role of decoherence and imperfect state preparation.

One direction to extend the results of our paper is to seek similar witnesses for discrete systems. One challenge in doing so is that the phase space approach for qubits is still not quite understand \cite{Raussendorf}. Thus, an intermediate task could be to start with odd-dimensional Hilbert space where a well-defined phase space formalism already exists \cite{Gibbons}. It is also interesting to consider more complicated encoding such as $x\rightarrow\rho^{n}(x)$ which leads to more extensive kernel $K(x,x')=\text{tr}(\rho^{n}(x)\rho^{n}(x'))$. This task requires generalizing (\ref{witnessW}) for $n+1$-points correlation functions.


{\it Acknowledgement.} I acknowledge discussions with Christoph Simon, Wilten Nicola, Reza Zamani, Hsin-Yuan Huang, Jarrod R. McClean and Barry Sanders. This work was supported by the Alberta Major Innovation Fund (MIF) Quantum Technologies project and by the Natural Sciences and Engineering Research Council (NSERC) of Canada.

\end{document}